\documentclass[12pt,a4paper]{article}%
\pdfoutput=1
\usepackage{jheppub}
\usepackage{amsmath,amssymb,amsfonts,wasysym}
\usepackage{mathrsfs}
\usepackage[bf,footnotesize]{caption2}
\usepackage{bbm}
\usepackage[]{hyperref}%
\hypersetup{
colorlinks=true,
linkcolor=red,
anchorcolor=black,
citecolor=blue,
filecolor=cyan,
menucolor=red,
runcolor=filecolor,
urlcolor=magenta,
bookmarksnumbered=true
}
\allowdisplaybreaks
\def\l{\label}

\def\({\left(}
\def\){\right)}
\def\f{\frac}
\def\be{\begin{equation}}
\def\ee{\end{equation}}
\def\bry{\begin{array}}
\def\ery{\end{array}}
\def\bes{\begin{subequations}}
\def\ees{\end{subequations}}
\def\bit{\begin{itemize}}
\def\eit{\end{itemize}}
\def\ben{\begin{enumerate}}
\def\een{\end{enumerate}}
\def\dst{\displaystyle}

\newcommand{\Dsl}{D\llap{/\kern+1.5pt}}
\newcommand{\MET}{E\llap{/\kern1.5pt}_T}

\def\nn{\nonumber}
\title{Liberating Higgs couplings in supersymmetry}
\author[a,b]{Christoffer Petersson}
\author[c,d,e]{Alberto Romagnoni}
\author[e,f,g]{Riccardo Torre}
\affiliation[a]{Physique Th\'eorique et Math\'ematique, 
Universit\'e Libre de Bruxelles, \\C.P. 231, 1050 Bruxelles, Belgium}
\affiliation[b]{International Solvay Institutes, Brussels, Belgium}
\affiliation[c]{The Mathematical Neuroscience Laboratory, CIRB / Coll\`ege de France (CNRS \\ UMR 7241, INSERM U1050, UPMC ED 158, MEMOLIFE PSL*), Paris, France}
\affiliation[d]{Departamento de F\'isica Te\'orica, Universidad Aut\'onoma de Madrid, \\ Cantoblanco, E-28049, Spain}
\affiliation[e]{Instituto de F\'isica Te\'orica UAM/CSIC,
Universidad Aut\'onoma de Madrid, \\ Cantoblanco, E-28049, Spain}
\affiliation[f]{Dipartimento di Fisica e Astronomia, Universit\'a di Padova and INFN Sezione di Padova, Via Marzolo 8, I-35131 Padova, Italy}
\affiliation[g]{SISSA, Via Bonomea 265, I-34136 Trieste, Italy}
\emailAdd{christoffer.petersson@ulb.ac.be}
\emailAdd{alberto.romagnoni@college-de-france.fr}
\emailAdd{riccardo.torre@cern.ch}
\proceeding{\vspace{-1.65cm}\flushright{\small DFPD-2012/TH/16\\
FTUAM-12-111\\
IFT-UAM/CSIC-12-103\\
LPN12-119\\}}
\abstract{
In the MSSM, the Higgs couplings to down-type quarks and leptons, normalized with respect to their corresponding Standard Model values, coincide at tree-level and this degeneracy is only slightly broken at the quantum level. Motivated by the latest results of the Higgs searches at the LHC and Tevatron, we explore the possibility of disentangling these couplings from each other by considering a scenario in which supersymmetry is broken spontaneously at a low scale. In such a scenario, all the Higgs couplings, except the ones to the $Z$ and $W$ bosons, receive tree level corrections that depend on the MSSM soft parameters. In particular, the corrections to the Yukawa couplings depend on the $A$-term soft parameter for the corresponding fermion, allowing for the freedom to break their usual relations, even in the MSSM decoupling limit. We highlight the main features of this scenario in terms of a benchmark point for which the normalized Higgs coupling to the tau leptons is depleted, the coupling to photons is enhanced, while all the other Higgs couplings, including the one to the bottom quarks, are close to their corresponding SM values. We also discuss the experimental bounds arising mainly from di-tau searches and comment on the discovery/exclusion prospects at the LHC.} 
\notoc
\begin{document}
\maketitle
\setcounter{page}{1}

\section{Introduction}

The ATLAS and CMS Collaborations at the LHC recently announced the discovery of a new boson, compatible with the Standard Model (SM) Higgs boson with a mass around 125 GeV \cite{ATLASCollaboration:2012tp,CMSCollaboration:2012ta}, in agreement with the hint of a signal seen by the CDF and D0 collaborations at the Tevatron \cite{CDFCollaboration:2012tx}. Even though there are at this stage no discrepancies in the data with sufficient statistical significance to disprove the SM Higgs compatibility, there are  at least two intriguing trends of possible deviation. The first one is the enhanced signal, with respect to the SM, in the $h\to \gamma\gamma$ decay mode, seen in both the ATLAS and CMS data \cite{ATLASCollaboration:2012tp,CMSCollaboration:2012ta}. For a partial list of models aiming at explaining this excess in terms of new physics, see Refs.~\cite{Hall:2011ty,Ellwanger:2011sv,Blum:2012wj,Boudjema:2012wpa,Bellazzini:2012tv,Azatov:2012ur,Hagiwara:2012ux,Benbrik:2012us,Buckley:2012vn,An:2012te,Alves:2012wp,Abe:2012vo,Bertolini:2012wg,ArkaniHamed:2012ur,Craig:2012tz,Batell:2012tc,Hashimoto:2012vw,Reece:2012ui,Alves:2012tb,Bertuzzo:2012uk,SchmidtHoberg:2012fw}. The second, less significant, possible deviation is a depleted signal in the $h\to \tau^+ \tau^-$ decay mode, seen by CMS \cite{CMSCollaboration:tw}. In this paper we discuss a supersymmetric (SUSY) scenario in which it is possible to accommodate a 125 GeV Higgs boson with couplings that can account for both of these deviations, while having the couplings to the other SM particles compatible with the corresponding SM values. 

In order to relate the measured Higgs signal rates to the Higgs couplings in a way that allows for a model-independent parametrization of the possible effects of new physics beyond the SM,  it is useful 
 to define the relevant partial decay widths in terms of Higgs couplings normalized with respect to the corresponding SM ones \cite{Giardino:2012vx,Montull:2012ve,Espinosa:2012wb,Carmi:2012wj,Group:2012te,Cacciapaglia:2012vq},
\be
\l{defcoupl}
\Gamma_{hii}   =  \left| c_i \right|^2 \Gamma_{hii}^{\mathrm{SM}}\,,
\ee
where $i=b,\tau,c,W,Z,g$ and $\gamma$. In terms of the dimensionless couplings $c_i$, the corresponding branching ratios (BRs) are given by, 
\begin{eqnarray}
\l{defBR}
\mathrm{BR}_{hii}   =  \left| \frac{c_i}{c_{\mathrm{tot}}} \right|^2 \mathrm{BR}_{hii}^{\mathrm{SM}}\,,\qquad\qquad \left| c_{\mathrm{tot}} \right|^2  =   \sum_i\left| c_i \right|^2 \mathrm{BR}_{hii}^{\mathrm{SM}}\,.
\end{eqnarray}
In $c_{\mathrm{tot}}$ we have neglected small contributions which can arise from, for example, $h\to\gamma Z$, as well as possible contributions from invisible decays that can be present in the scenarios we will consider, such as the Higgs decays into two goldstinos (see, for example, Ref.~\cite{Antoniadis:2010hs}), or other non-standard decays, such as the Higgs decay into a goldstino and a neutralino, which subsequently decays into a goldstino and photon, giving rise to a  final state with a monophoton and missing transverse energy \cite{Petersson:2012tl}. In Eqs.~\eqref{defcoupl} and \eqref{defBR}, the SM values are recovered upon setting $c_i=1$. Note that, in contrast to other definitions appearing in the literature, we have also normalized $c_g$ and $c_\gamma$ with respect to their corresponding SM  couplings. 

The signal rates in the $h\to ZZ^{\ast}$ and $h\to WW^\ast$ channels seem to be in good agreement with the SM rates \cite{ATLASCollaboration:2012tp,CMSCollaboration:2012ta}. Therefore, unless there are some compensating effects between the relevant production cross sections, the total width and the individual partial widths, the simplest explanation of this agreement is that the most relevant couplings, namely $c_g$, $c_Z$, $c_W$ and $c_b$, are all close to one, i.e.~their corresponding SM value. If these four couplings are close to one, then the only way to significantly enhance the $h\to \gamma\gamma$ signal is to invoke some new physics that only affects $c_\gamma$. For example, in the context of the minimal supersymmetric extension of the SM (MSSM), it is possible to increase the $c_\gamma$ coupling without affecting the other couplings by considering the 1-loop contribution to $c_\gamma$ arising from stau sleptons \cite{Carena:2011uq,Carena:2012uq,Carena:2012wl}.

For a value of $c_b$ close to one, within the parameter space of the MSSM, it is difficult to accommodate a value of $c_\tau$ that is significantly smaller than one. The reason is that the same (down-type) Higgs doublet enters in both the bottom and tau Yukawa couplings. Hence the tree level values of $c_b$ and $c_\tau$ are the same, given by $-\sin\alpha/\cos\beta$, where $\alpha$ is the mixing angle.  In fact, for small values of $\tan\beta$, $c_b$ and $c_\tau$ are almost identical even at the quantum level. For large values of $\tan\beta$ it was estimated in Refs.~\cite{Gupta:2012uh,Peskin:2012tm}, using Refs.~\cite{Carena:1999cy,Haber:2006jz}, that, due to $\tan\beta$-enhanced loop effects, $c_b$ may differ from $c_\tau$ by a few percent at the quantum level. Nevertheless, the conclusion is that, in the MSSM, $c_b$ and $c_\tau$ are expected to be closely related in all of the parameter space. 

In the MSSM literature it is in general assumed that the scale of SUSY breaking $\sqrt{f}$, or equivalently the gravitino mass\footnote{In flat spacetime, $\sqrt{f}$ and the gravitino mass $m_{3/2}$ are related by $m_{3/2}=f/(\sqrt{3}M_{\mathrm{P}})$, where $M_{\mathrm{P}}$ is the Planck mass.}, is sufficiently large in order for the degrees of freedom associated with the spontaneous breaking of SUSY not to be relevant for the low energy effective theory. These degrees of freedom involve the Goldstone fermion, the goldstino \cite{1977PhLB...70..461F,1988PhLB..215..313C}, and if SUSY is linearly realized, also the complex scalar superpartner of the goldstino, the sgoldstino \cite{Brignole:1997hb,Perazzi:2000ku,Perazzi:2000dk,Gorbunov:2000ii,Gorbunov:2002co,Brignole:2003hb,Demidov:2004uy,Bertolini:2011wj,Petersson:2011in,Antoniadis:2012ui}. The goldstino and sgoldstino interactions with the MSSM fields are dictated by supercurrent conservation and the strength of the interactions are determined in terms of ratios of the MSSM soft parameters over $f$. Since the sgoldstino is not protected by the Goldstone shift symmetry, it generally acquires a mass upon integrating out some heavy fields of the hidden sector, but the precise value of its mass is model-dependent. In this paper we consider $\sqrt{f}$ to be around 5 TeV and the sgoldstino mass to be in the range between 500 GeV and 1.5 TeV. In Ref.~\cite{Bellazzini:2012ul} it was shown that in such a scenario, via a small sgoldstino mixing with a SM-like Higgs scalar $h$, it is possible to increase the $c_\gamma$-coupling for $h$ without significantly affecting any of the other $h$ couplings. 

In this work we extend the analysis of Ref.~\cite{Bellazzini:2012ul} by also considering the effect of the sgoldstino mixing on the  Higgs couplings to the fermions. The interactions between the sgoldstino and the fermions arise through supersymmetric operators which also give rise to the $A$-terms.  Therefore, the strength of these interactions will depend on ratios of the corresponding $A$-term parameter over $f$. 
Due to this $A$-term dependence, even in the MSSM decoupling limit, it is possible to significantly modify the Higgs couplings to fermions. The fact that each Higgs coupling to fermions depend on its individual $A$-term parameter allows for the possibility of, for example, breaking the otherwise degenerate tree-level values of the tau and bottom Higgs couplings.

This paper is organized as follows: In Section \ref{Sgoldcont} we discuss the SUSY operators that give rise to the relevant sgoldstino interactions and we show how the sgoldstino mixing contributions are incorporated in the Higgs couplings. In Section \ref{parameters}, in order to illustrate the particular features of the scenario  under consideration, we discuss a benchmark model which accommodates a Higgs particle $h$ with a mass at around 125 GeV, for which the $h\to \gamma\gamma$ signal is enhanced,  the $h\to \tau^+ \tau^-$ signal is depleted, while the signal rates in all the other channels are close to their corresponding SM values. We also discuss experimental bounds on the sgoldstino-like scalar state, arising mainly from di-tau searches, and its discovery potential at the LHC. In Section \ref{conclusion} we summarize and conclude.

\section{Tree-level mixing contributions to the Higgs couplings} 
\l{Sgoldcont}
 
In the framework we consider, SUSY is linearly realized but spontaneously broken. In such a framework, the soft terms are promoted to supersymmetric operators involving a goldstino superfield $X=x+\sqrt{2}\theta G+\theta^2 F_X$, where $G$ is the goldstino \cite{1977PhLB...70..461F,1988PhLB..215..313C}, $F_X$ is the auxiliary component that acquires a non-vanishing vacuum expectation value (VEV) $f$, and $x$ is the scalar superpartner of the goldstino, the sgoldstino \cite{Brignole:1997hb,Brignole:2003hb,Bertolini:2011wj,Petersson:2011in,Perazzi:2000ku,Perazzi:2000dk,Gorbunov:2000ii,Gorbunov:2002co,Demidov:2004uy,Antoniadis:2012ui}. 
In the limit where the sgoldstino mass is large compared to the other mass scales in the theory, the sgoldstino can be integrated out and in the low energy effective theory, SUSY is non-linearly realized, see, e.g., Ref.~\cite{Komargodski:2009cq,Antoniadis:2010hs,Dudas:2011wk,Antoniadis:2011ve,Farakos:2012tk}. In this paper we consider the case where the sgoldstino mass is of the same order as the MSSM superparters masses and therefore, the sgoldstino is treated dynamically.

In addition to the usual SUSY operators in the MSSM, we also consider the following SUSY operators, which give rise to the Higgs soft masses\footnote{Similar operators which give rise to the sfermion soft masses are of course also present.}, the sgoldstino soft mass, a non-vanishing vacuum energy, the $B_\mu$-term, the trilinear scalar interactions between the sgoldstino and the Higgs scalars, and the gaugino masses, 
\begin{align}
& -\int d^4 \theta \left( \sum_{I=d,u} \frac{m_{I}^2}{f^2}   \, X^\dagger X H_{I}^{\dagger} e^{\,gV} H_{I} +\frac{m_x^2}{4f^2} \left(X^\dagger X\right)^2 \right)~, \l{LkinX}\\
& \int d^2 \theta \left(fX - \frac{B_\mu}{f} X H_d \,H_u - \frac{A_x}{f} X^2 H_d \,H_u \right)  +\mathrm{h.c.}\,, \l{LBmuX} \\
& -\sum_{i=1}^{3}\,\frac{m_{i}}{2f}\int d^2\theta \, X \,W_{A_i}^\alpha W_{\alpha}^{A_i}+ \mathrm{h.c.} \,,\l{susyop1}
\end{align}
where the indices $A_1=1$, $A_2=1,2,3$ and $A_3=1,\cdots,8$, see Ref.~\cite{Petersson:2011in} for details.\footnote{If instead $X$ had been a non-linear goldstino superfield and satisfied the quadratic constraint of Ref.~\cite{Komargodski:2009cq}, the $m_x$-operator in Eq.~\eqref{LkinX} and the $A_x$-operator in Eq.~\eqref{LBmuX} would not have been present.} 
The MSSM $A$-terms can be generated by the following SUSY operators,
\be\l{susyop3}
-\int d^2\theta   \( \frac{A_u}{f}X\,Q \,H_u\, U^c+\frac{A_d}{f} X\,Q \,H_d\, D^c+\frac{A_l}{f} X\, L \,H_d\, E^c\)+\mathrm{h.c.}\,,
\ee
where the flavor indices are suppressed and we assume $A_u, A_d$ and $A_l$ to be proportional to the corresponding SM Yukawa matrices, with free proportionality coefficients. In order for the operator expansion in terms of higher dimensional supersymmetric operators to be perturbative, we require that the (absolute value of the) soft parameters $m_d^2, m_u^2, m_{x}^2, A_{x}^2, m_i^2, A_u^2, A_d^2, A_l^2, B_{\mu}< f$. Also, we take all the parameters to be real.

In addition to the trilinear couplings between the Higgs scalars and the sfermions, arising from when the auxiliary component of $X$ acquires a VEV, the operators in Eq.~\eqref{susyop3} also generate Yukawa-like interactions between the sgoldstino $x$ and the SM fermions, arising from when the Higgs scalars acquire VEVs. Upon diagonalization of the neutral CP-even scalar mass matrix, these sgoldstino interactions give rise to extra contributions to the Yukawa couplings of the lightest physical Higgs scalar $h$, without affecting the corresponding fermion masses. 
By taking into account this kind of sgoldstino mixing effects, the dimensionless $h$ couplings in Eq.~\eqref{defcoupl} will have the following structure, 
\begin{align}
& \dst c_b = \frac{R_{(h,d)}}{\cos\beta}+\frac{A_b \,v^2 \cos\beta\,R_{(h,x)}}{\sqrt{2}\,m_b\,f}\,, 
\nn \\
&\dst c_{\tau} = \frac{R_{(h,d)}}{\cos\beta}+\frac{A_\tau \,v^2 \cos\beta\,R_{(h,x)} }{\sqrt{2}\,m_\tau \,f}\,,
\nn \\
&\dst c_c  =  \frac{R_{(h,u)}}{\sin\beta}+\frac{A_c \,v^2 \sin\beta\,R_{(h,x)}}{\sqrt{2}\,m_c\,f}\,, 
 \nn \\
& \dst c_t  = \frac{R_{(h,u)}}{\sin\beta}+\frac{A_t \,v^2 \sin\beta\,R_{(h,x)}}{\sqrt{2}\,m_t\,f}\,, \l{ct}
 \\
& \dst c_V = c_W=c_Z=\sin\beta \,R_{(h,u)}+\cos\beta \,R_{(h,d)}\,, 
\nn\\
&\dst c_g  = \frac{c_t \mathcal{A}_t + c_b \mathcal{A}_b +\frac{m_{3}}{2\sqrt{2}\,f} \frac{12\pi\,v}{\alpha_s} R_{(h,x)}}{\mathcal{A}_t +  \mathcal{A}_b}\,,
\nn \\
&\dst c_\gamma  =  \frac{\frac{2}{9}c_t \mathcal{A}_t - \frac{7}{8}c_V \mathcal{A}_W +
 \frac{m_{1} \cos^{2} \theta_W +m_{2} \sin^{2} \theta_W }{2\sqrt{2}\, f } \frac{\pi\, v }{\alpha} R_{(h,x)}}{\frac{2}{9} \mathcal{A}_t - \frac{7}{8} \mathcal{A}_W}\,, \nn 
\end{align}
where the 1-loop form factors, for a Higgs mass around 125 GeV, are given by $\mathcal{A}_t\approx1.03$, $\mathcal{A}_b\approx-0.06+0.09i$ and $\mathcal{A}_W\approx1.19$ \cite{Carmi:2012wj}. The matrix elements $R_{(h,u)}, R_{(h,d)}$ and $R_{(h,x)}$ of the rotation matrix $R$, which diagonalizes the neutral CP-even scalar mass matrix, are the mixing coefficients of the lightest physical Higgs scalar $h$ with the gauge basis scalar fields $h_u, h_d$ and $x$, respectively. The sgoldstino mixing contributions to the couplings of $h$ in Eq.~\eqref{ct} are always proportional to $R_{(h,x)}/f$, implying that they are negligible when either the SUSY breaking scale is too large or when the mixing with the sgoldstino scalar is too small. 

The usual MSSM Higgs couplings can be recovered by setting the mixing element $R_{(h,x)}^{\mathrm{MSSM}}=0$, $R_{(h,u)}^{\mathrm{MSSM}}=\cos\alpha$ and  $R_{(h,d)}^{\mathrm{MSSM}}=-\sin\alpha$. The SM Higgs couplings are then recovered by taking the decoupling limit, where Higgs pseudoscalar mass $m_A$ is far greater than the $Z$-boson mass $m_Z$, such that $\alpha\to\beta-\pi/2$, $\cos\alpha\to \sin\beta$, $\sin\alpha\to -\cos\beta$ and the dimensionless couplings in Eq.~\eqref{ct} become $c_i\to c_i^{\mathrm{SM}}=1$. In this class of models, $c_V=c_Z=c_W$ in Eq.~\eqref{ct} is unchanged with respect to the MSSM formula, and upon replacing $R_{(h,u)}$ and $R_{(h,d)}$ with their MSSM expressions, we obtain the usual expression $\sin(\beta-\alpha)$, which approaches one in the MSSM decoupling limit.

The sgoldstino mixing contributions to the Higgs couplings to the fermions in Eq.~\eqref{ct} depend on the $A$-term parameters and are suppressed by the mass of the  corresponding fermion. This implies that it would require a very large top $A$-parameter in order to even have a slight effect on the $c_t$ coupling in Eq.~\eqref{ct}. Instead, it is easier to affect the bottom and tau Higgs couplings by choosing appropriately the free proportionality coefficients between the bottom and tau $A$-terms and the corresponding Yukawas. This allows for, e.g., a  splitting between the $c_b$ and $c_\tau$ couplings in  Eq.~\eqref{ct}. Such a tree level splitting of the bottom and tau Higgs couplings is not possible in the usual MSSM setup. 

Let us now define the Higgs production cross sections and signal rates in terms of the $c_i$-couplings in Eq.~\eqref{ct}. Instead of fitting the precise values of the measured signal rates, we will consider the inclusive channels, dominated by the gluon-gluon fusion production mode, except for the $h\to b\bar{b}$ decay mode for which we consider the vector boson associate production mode, since it is more relevant for the LHC searches. Note that, even though the depletion seen for $h\to \tau^+ \tau^-$ in  the current CMS data concerns the vector boson fusion production mode, we will only consider the inclusive $h\to \tau^+ \tau^-$ channel. This is partly due to the fact that we do not know the precise selection efficiencies for the various production modes contributing to the di-jet final state category of this channel. But mainly, as already mentioned, this is because our aim is not to fit any precise signal strengths but rather to show that it is possible, within the framework of low scale SUSY breaking with a sgoldstino scalar, to significantly separate the $c_\tau$ and $c_b$ couplings in  Eq.~\eqref{ct}.   

The gluon-gluon fusion (ggF), vector boson fusion (VBF) and vector boson associated (VH) production  cross sections, normalized to their SM values, are given \mbox{by \cite{Carmi:2012wj}},
\be
\frac{\sigma_{\mathrm{ggF}}}{\sigma_{\mathrm{ggF}}^{\mathrm{SM}}}=\left| c_g \right|^2~~,~~
\frac{\sigma_{\mathrm{VBF}}}{\sigma_{\mathrm{VBF}}^{\mathrm{SM}}}=\frac{\sigma_{\mathrm{VH}}}{\sigma_{\mathrm{VH}}^{\mathrm{SM}}}=\left| c_V \right|^2~.
\ee
 The normalized total inclusive production cross section is therefore given by,
\be
\frac{\sigma_{\mathrm{tot}}}{\sigma_{\mathrm{tot}}^{\mathrm{SM}}}= \frac{\left| c_g \right|^2\sigma_{\mathrm{ggF}}^{\mathrm{SM}}+\left| c_V  \right|^2 \left( \sigma_{\mathrm{VBF}}^{\mathrm{SM}}+\sigma_{\mathrm{VH}}^{\mathrm{SM}}\right)}{\sigma_{\mathrm{ggF}}^{\mathrm{SM}}+\sigma_{\mathrm{VBF}}^{\mathrm{SM}}+\sigma_{\mathrm{VH}}^{\mathrm{SM}}}~
\ee
and the signal rates in the inclusive channels can be written as,
\be\l{incl}
R^{\mathrm{incl}}_{ii} = \frac{\sigma_{\mathrm{tot}}}{\sigma_{\mathrm{tot}}^{\mathrm{SM}}}\frac{\mathrm{BR}_{hii} }{\mathrm{BR}_{hii}^{\mathrm{SM}}}\,,
\ee
for which we will only consider the decay modes $i=Z,W,\gamma$ and $\tau$.  
Finally, the signal rate for the associated production of $b\bar{b}$ via VH can be written as,
\begin{eqnarray}
R^{\mathrm{VH}}_{bb} & = & \frac{\sigma_{\mathrm{VH}}}{\sigma_{\mathrm{VH}}^{\mathrm{SM}}}\frac{\mathrm{BR}_{hbb} }{\mathrm{BR}_{hbb}^{\mathrm{SM}}}~.
\end{eqnarray}

\section{Numerical examples and phenomenology} 
\l{parameters}

In this Section we illustrate the main features of the scenario under consideration by discussing a benchmark point in the parameter space. In order to highlight the features that are specific to this scenario  we choose extreme values of the key parameters. Some of these extreme values are chosen such that the resulting spectrum contains a sgoldstino-like scalar that is on the verge of being excluded/discovered by di-tau searches at the LHC.  Together with the requirement of having all the soft parameters smaller than the SUSY breaking scale, the extreme values of the parameters are meant to show, for example, how large is the splitting between the $c_\tau$ and $c_b$ couplings, and their corresponding signal rates, that can be accommodated. Clearly, less extreme parameter values give rise to smaller effects and we discuss how the couplings and signal rates vary as we move in the parameter space.

\subsection{A benchmark point} 

The values of the parameters in the benchmark point we consider are shown in the left column of Table \ref{bench} and the Higgs mass, $c$-couplings and signal rates these values of the parameters give rise to, are shown in the right column of Table \ref{bench}. Let us outline the logic behind this choice of parameters.  
\begin{table}[t]
\begin{center}
\begin{tabular}{l|r||c||l|r}
{\bf Parameter} & {\bf Value}	&\quad\quad& {\bf Quantity} 	& {\bf Value}\\ \hline
$\sqrt{f}$ 		& $5$ TeV 	&& $m_{h}$ 				& $125.8$ GeV\\
$m_{12}$		&  $-1.5$ TeV	&& $c_{g}$ 				& $1.1$\\
$m_{3}$		& $900$ GeV	&& $c_{\gamma}$ 			& $1.29$\\
$m_{x}$		& $650$ GeV	&& $c_{V}$ 				& $0.99$\\
$m_{\tilde{t}}$	& $370$ GeV	&& $c_{t}$ 				& $0.99$\\
$\tan\beta$	&  $ 1$		&& $c_{b}$ 				& $1.13$\\
$\mu$		&  $ 100$ GeV	&& $c_{\tau}$ 				& $0.68$\\
$B_{\mu}/f$	&  $0.7$		&& $c_{\text{tot}}^{2}$ 		& $1.3$\\
$A_{x}$		&  $ 90$ GeV	&&&\\
$A_{t}$		& $0$ GeV	&& $R_{ZZ,WW}^{\text{incl}}$ 	& $1.03$\\
$A_{b}$		& $4$ TeV		&& $R_{\gamma\gamma}^{\text{incl}}$ & $1.74$\\
$A_{c}$		& $0$ GeV	&& $R_{bb}^{VH}$ 			& $1.1$\\
$A_{\tau}$		& $-4$ TeV	&& $R_{\tau\tau}^{\text{incl}}$ 	& $0.48$\\
\hline
\end{tabular}
\end{center}
\caption{\small Values of the parameters in the benchmark point and resulting Higgs mass, $c$-couplings and signal rates.}
\l{bench}
\end{table}

In an attempt to isolate the effects that are specific to this scenario we consider only the case in which the mixing between the two Higgs doublets are negligible, corresponding to taking the MSSM decoupling limit, $m_A^2\gg m_Z^2$. 
This can be accomplished by considering a large value of $B_\mu$. In scenarios where the SUSY breaking scale is low, due to presence of $F$-term contributions depending on ratios of soft parameters over the SUSY breaking scale, the tree-level mass of the lightest neutral CP-even Higgs scalar can be significantly larger than the corresponding MSSM value \cite{Antoniadis:2010hs,Petersson:2011in}. 

From the analysis done in Ref.~\cite{Petersson:2011in} it can be seen that for large values of $B_\mu$, the tree level Higgs mass is maximized for small values of $\tan\beta$ and $\mu$.\footnote{Moreover, as it is evident from the Eqs.~\eqref{ct}, the sgoldstino mixing contributions to the $c_b$ and $c_\tau$ couplings, in the decoupling limit where $R_{(h,d)}= \cos \beta$, are maximized for low values of $\tan \beta$. } In order to satisfy the experimental bound on the charginos, we set $\mu$ to be 100 GeV. In the regime where $f, B_\mu \gg m^2_x \gg v^2, \mu^2, A_x^2$ the dominant contributions  to the mass of the lightest Higgs scalar are given by,
\be\l{mh}
m_h^{2} = m_{Z}^2 \cos^2 2\beta+ v^2 \(  \frac{B_{\mu}^2}{2 f^2} \sin^2 2\beta -\frac{2 A_x^2  \sin ^2 2 \beta }{m_x^2}+ \delta \)\,,
\ee
where the 1-loop contribution is given by
\begin{eqnarray} \l{delta}
&&\delta= \frac{3 m_t^4}{2 \pi^2 v^4} \left[ \log \(\frac{m_{\tilde{t}}^2}{m_t^2} \)+  \frac{X_t^2}{m_{\tilde{t}}^2} \(1 - \frac{X_t^2}{12 m_{\tilde{t}}^2} \) \right]\,,
\end{eqnarray}
with $X_t = A_t - \mu/\tan \beta$. Note that the ratio $ B_{\mu}/f$ in Eq.~\eqref{LBmuX} plays the analogous role of the parameter $\lambda$ in the context of the NMSSM, see Ref.~\cite{Ellwanger:2009en} for a review and references. Inspired by the commonly used value of $\lambda$, in the benchmark point we set $ B_{\mu}/f=0.7$. 

Due to the presence of the extra tree-level contributions with respect to the usual MSSM contribution, corresponding to the first term in Eq.~\eqref{mh}, the 1-loop correction in Eq.~\eqref{delta} is not as important as in the MSSM. In order to highlight this feature of the scenario, we consider small 1-loop corrections in which $A_t=0$ (implying that the whole $A_u$ matrix in Eq.~\eqref{susyop3} is vanishing and in particular $A_c=0$ in Eq.~\eqref{ct}) and the stop mass is 370 GeV, where the latter specific value is chosen such that the Higgs mass is around 125 GeV. 
Note that the dependence on the value of the $A_t$-parameter of sgoldstino mixing contribution to the $c_t$-coupling in Eq.~\eqref{ct} is negligible due to the suppression by the top quark mass. Therefore, apart from the freedom in raising the Higgs mass via loop corrections, our results are almost insensitive to its value.

In Eq.~\eqref{mh} we see that the tree-level mixing induced by $A_x$ always acts in a destructive way. This is due to the the fact that the $h$ scalar corresponds to the smallest eigenvalue of the mass matrix and therefore, due to level repulsion, any mixing with the heavier sgoldstino-like $\phi$ scalar will decrease $m_h$.

The $c_g$ coupling in Eq.~\eqref{ct} depends on the gluino mass, which is experimentally constrained from below. As discussed in Ref.~\cite{Bellazzini:2012ul}, in order for the sgoldstino-like scalar  to evade bounds from direct di-photon and di-jet searches, we choose \mbox{$\sqrt{f}=5$ TeV}. This choice is consistent with what we find in Section \ref{sgoldbound} concerning bounds arising from di-tau searches.  Moreover, in order not to have a too large deviation, with respect to the SM, in the gluon-gluon fusion production cross section, since $c_g$ depends on the combination $m_3 R_{(h,x)}/f$, even if we set the gluino mass at it the experimental bound at around 900 GeV, we are required to have a small value for $R_{(h,x)}$, around 0.1. 

In the regime of the parameter space under consideration, the mixing elements can be approximated by,
\be\l{mix}
\bry{l}
\dst R_{(h,x)}=  \frac{3 \sqrt{2} A_x v \sin 2\beta}{2 m_x^2}\,, \\
\dst R_{(h,d)}= \cos \beta -\frac{A_x^2 v^2 \sin^2\beta}{2 m_x^4} \left(14 \cos \beta+ 5 \cos 3\beta \right)\,,   \\
\dst R_{(h,u)}= \sin \beta -\frac{A_x^2 v^2 \cos^2\beta}{2 m_x^4} \left(14 \sin \beta - 5 \sin 3\beta \right)\,. 
\ery
\ee
For $A_x \ll m_x$, the $R_{(h,d)}$ and $R_{(h,u)}$ elements are close to their corresponding values in the MSSM decoupling limit. In the benchmark point, $A_{x}$ and $m_x$ are chosen in order to have the desired value for $R_{(h,x)} $, while having $m_x$ large enough to evade the bounds on the sgoldstino-like scalar.

All the other coefficients, namely $m_1, m_2, A_b, A_c$ and $A_\tau$, do not directly affect the Higgs sector, but only the $c$-couplings in Eq.~\eqref{ct}, and they are chosen in order to have a significant splitting of $c_b$ and $c_\tau$, an increase in the $c_\gamma$-coupling while keeping the other couplings close to their corresponding SM values. Concerning the $A_b$, we could take it to be vanishing in order to have the $c_b$-coupling close to its SM value. However, the $c_b$-coupling is the most relevant one for the total width and, in the benchmark point, the gluon-gluon fusion production cross section is larger than in the SM. Therefore, if we want the inclusive signal rates for the $h\to ZZ^\ast$ and $h\to WW^\ast$ decay modes to be as close to the SM values as possible, we can use $A_b$ in order to increase $c_b$. In this way, we increase the total width and hence make the inclusive $h\to ZZ^\ast$ and $h\to WW^\ast$ channels more compatible with the SM values. Of course, it is also possible to use $A_b$ in order to decrease $c_b$ and reduce the partial width of the $h\to b\bar{b}$ decay. The values of the couplings in Eq.~\eqref{ct} and the signal rates this benchmark point gives rise to, are shown in the right column of Table \ref{bench}.

\subsection{The parameter space}

Let us now discuss how the $c$-couplings and signal rates vary as we move away, in the parameter space, from the benchmark point in Table \ref{bench}. Note that we will at all times only consider the MSSM decoupling limit.

In the left panel of in Fig.~\ref{fig:gauginos}, the relevant quantities are shown as a function of the linear combination $m_{12} = m_{1} \cos^{2} \theta_W +m_{2} \sin^{2} \theta_W $. We see that only the rate $R^{\text{incl}}_{\gamma \gamma}$ varies, while all other quantities are independent of this linear combination of bino and wino masses. In right panel of in Fig.~\ref{fig:gauginos} we see that all the rates of the inclusive channels increase as we increase the value of the gluino mass, since it enters in the \mbox{$c_g$-coupling} in Eq.~\eqref{ct} and therefore in the gluon-gluon fusion production cross section, which is the dominant production mode in the inclusive channel. In contrast, for the rate  $R^{\mathrm{VH}}_{b b}$ the $c_g$ coupling does not enter in the production cross section, but only in terms of a contribution to the total width. Note that by choosing the opposite sign for the gluino mass, with respect to our benchmark point, i.e.~$m_3 = - 900$ GeV, implies that, for example, the rate $R^{\mathrm{incl}}_{ZZ}=R^{\mathrm{incl}}_{WW}$ is well below the SM value. Since we can not raise the $c_Z=c_W$ above one in Eq.~\eqref{ct} the only possibility we have in order to bring this rate closer to the SM value is to decrease the total width, in particular by decreasing $c_b$ in Eq.~\eqref{ct}. 

\begin{figure*}[!t]
\begin{center}
\hspace{-0.35cm}\includegraphics[scale=0.35]{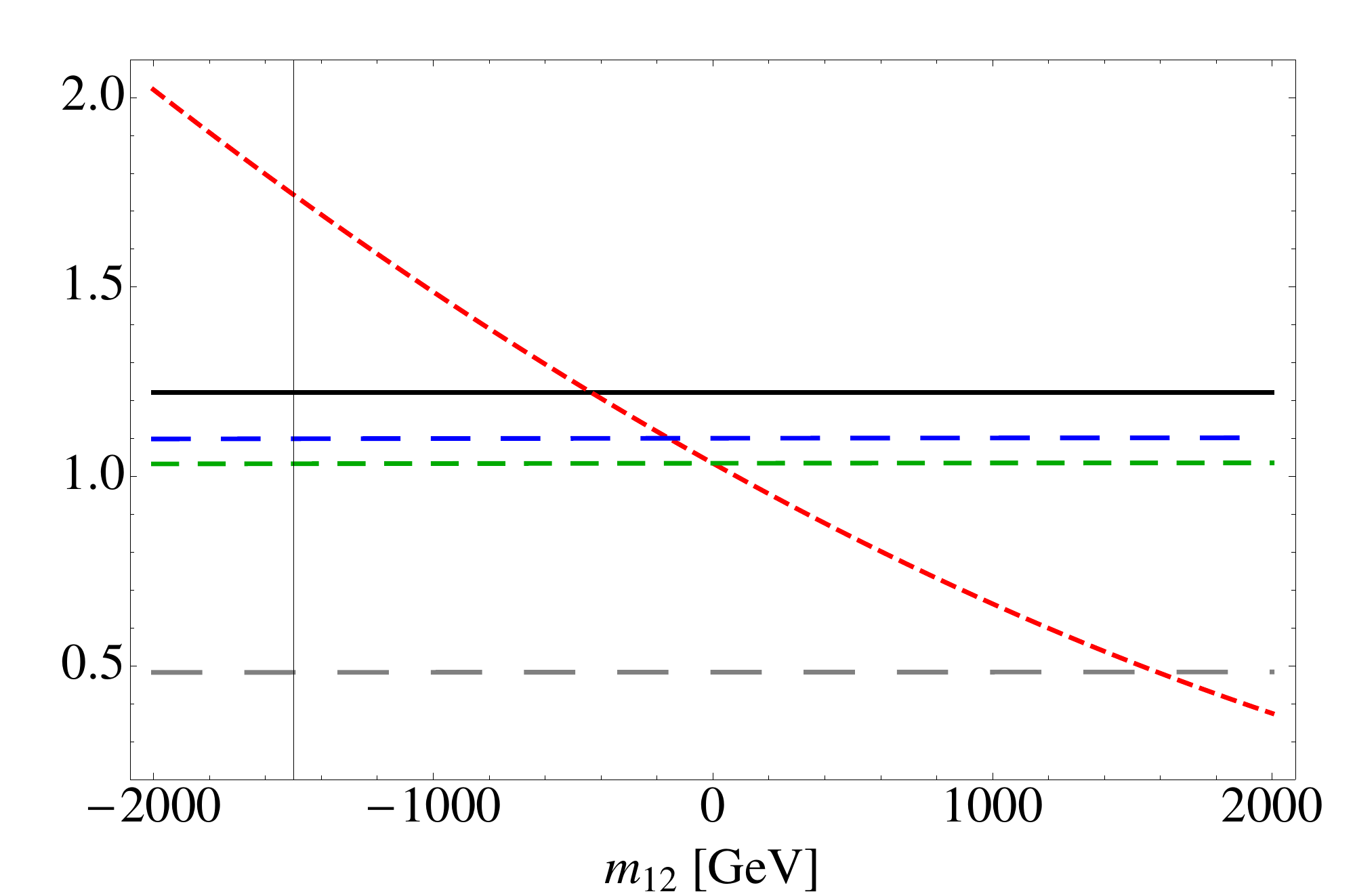}\hspace{-0.4cm}
\includegraphics[scale=0.37]{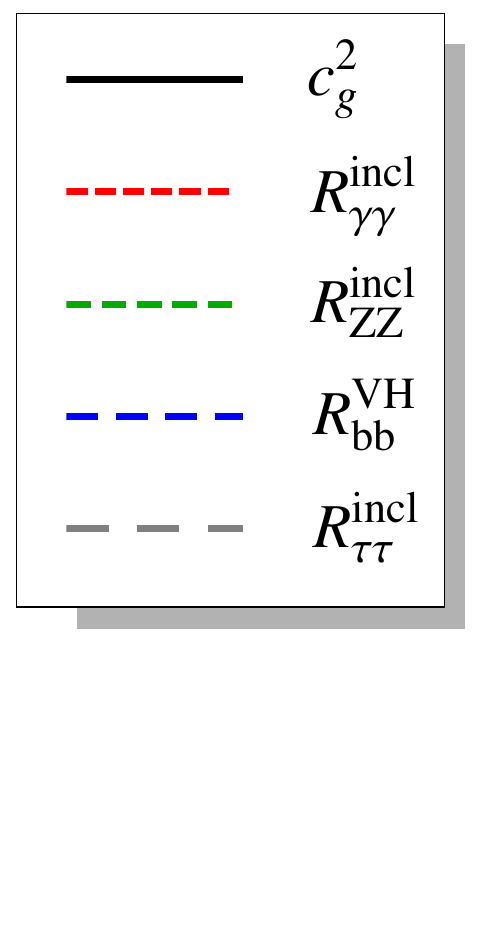}\hspace{-0.45cm}
\includegraphics[scale=0.35]{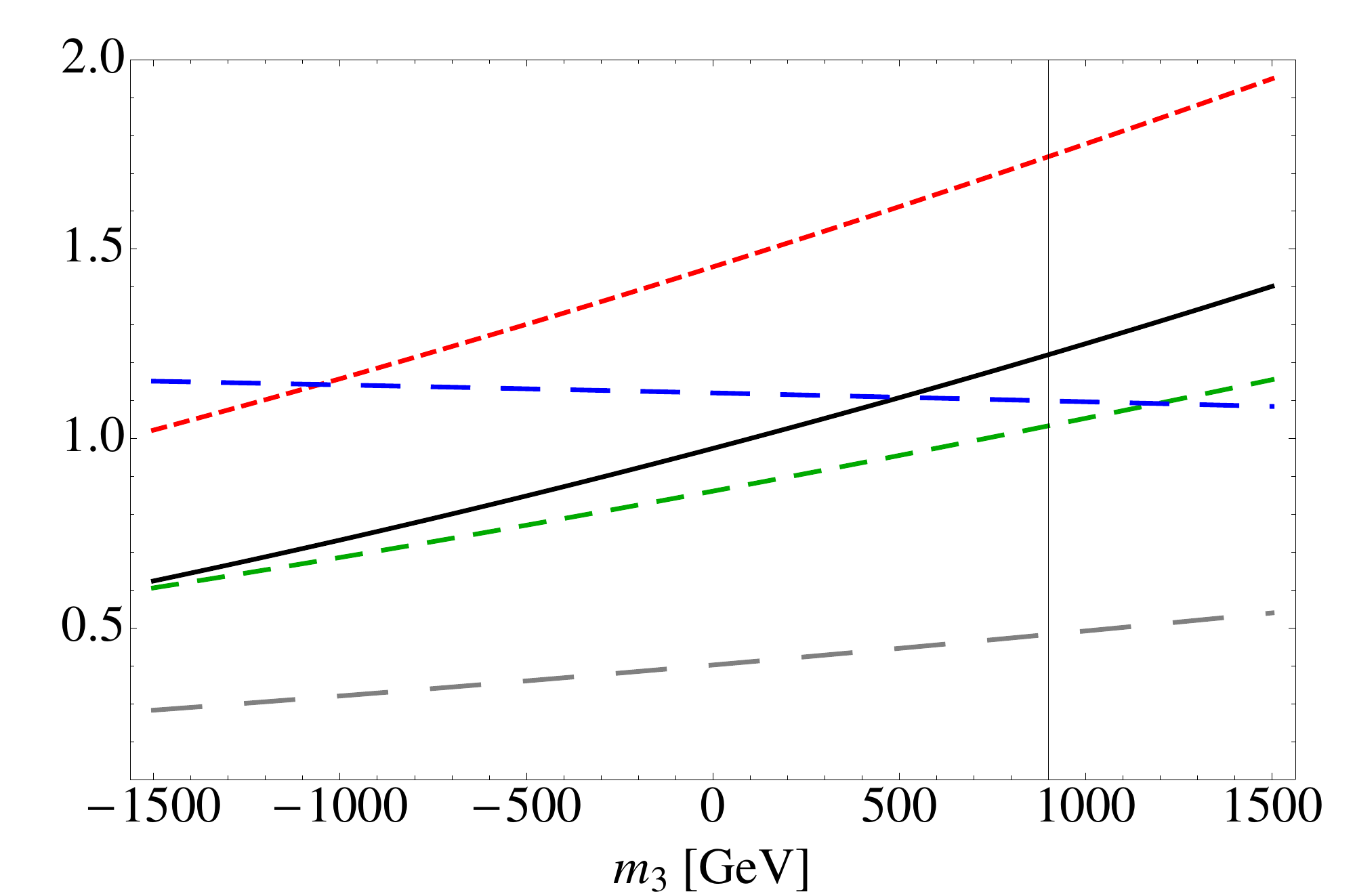}
\end{center}
\caption{ 
\small The signal rates and the coefficient $c_g^2$, defined in the text, as functions of the combination of the gaugino masses $m_{12} = m_{1} \cos^{2} \theta_W +m_{2} \sin^{2} \theta_W $  (left panel) and of the gluino mass $m_3$ (right panel). In each case we keep fixed all the other parameters at their values given in Table \ref{bench}. The vertical lines correspond to the benchmark point values.
}\l{fig:gauginos}
\end{figure*}

The plots in Fig.~\ref{fig:Aterm} show the dependence of  the relevant quantities on the $A$-term parameters $A_b$ and $A_\tau$. We see that, due to the fact that the SM Higgs coupling to bottom quarks is larger than the coupling to tau leptons, $R^{\mathrm{incl}}_{\tau\tau}$ has a stronger dependence on $A_\tau$ than $R^{\mathrm{VH}}_{bb}$ has on $A_b$. We also notice that, since $c_b$ is more relevant than $c_\tau$ for the total width, as $A_b$ is increased, the total width is increased, which reduces all the BRs, except the one for $h\to b\bar{b}$. This is even more evident in Fig.~\ref{fig:contour}, where the dependence of the signal rates on $A_b$ and $A_\tau$ are shown. We see how the contours for $R^{\mathrm{incl}}_{ZZ}=R^{\mathrm{incl}}_{WW}$ and those of $R^{\mathrm{incl}}_{\gamma\gamma}$ are more or less insensitive to the value of $A_\tau$ whereas, in contrast, the ratio $R^{\mathrm{incl}}_{\tau\tau}/R^{\mathrm{VH}}_{bb}$ is  strongly dependent.

All the relevant quantities obviously  depend strongly on the value of the SUSY breaking scale $\sqrt{f}$. In Fig.~\ref{fig:fdep} it is shown how the sgoldstino mixing contributions in Eqs.~\eqref{ct} decrease for increasing values of $\sqrt{f}$. The thin lines in Fig.~\ref{fig:fdep} correspond to the choice $A_b=-A_\tau=4$ TeV while the thick lines corresponds to the limiting case in which $A_b=-A_\tau=\sqrt{f}$. In the latter case, significant effects can be achieved even for a large value of $\sqrt{f}$. Obviously, the larger the value for the $A$-terms, the larger the corresponding soft mass has to be, in order to avoid  tachyonic sfermions. 

As a final comment, due to our assumptions concerning the flavor structure of the $A$-terms, we do not expect any significant contributions to flavor changing processes. Note that new diagrams are present, analogous to the SM Higgs contributions to processes such as $b\to s\gamma$ and $\tau \to l\gamma$, with the sgoldstino replacing the Higgs. However, these diagrams turn out to be negligible since the Higgs contribution is already subleading with respect to the one from the gauge bosons, and the sgoldstino contribution is suppressed with respect to the Higgs one by the factor \mbox{$g_{\phi \bar{f}f}^{2}/g_{h \bar{f}f}^{2}\cdot m_{h}^{4}/m_{\phi}^{4}$}, where $g_{\phi \bar{f}f}^{2}$ and $g_{h \bar{f}f}^{2}$ are, respectively, the couplings of the sgoldstino and the Higgs to two fermions.

\begin{figure*}[!t]
\begin{center}
\hspace{-0.35cm}\includegraphics[scale=0.35]{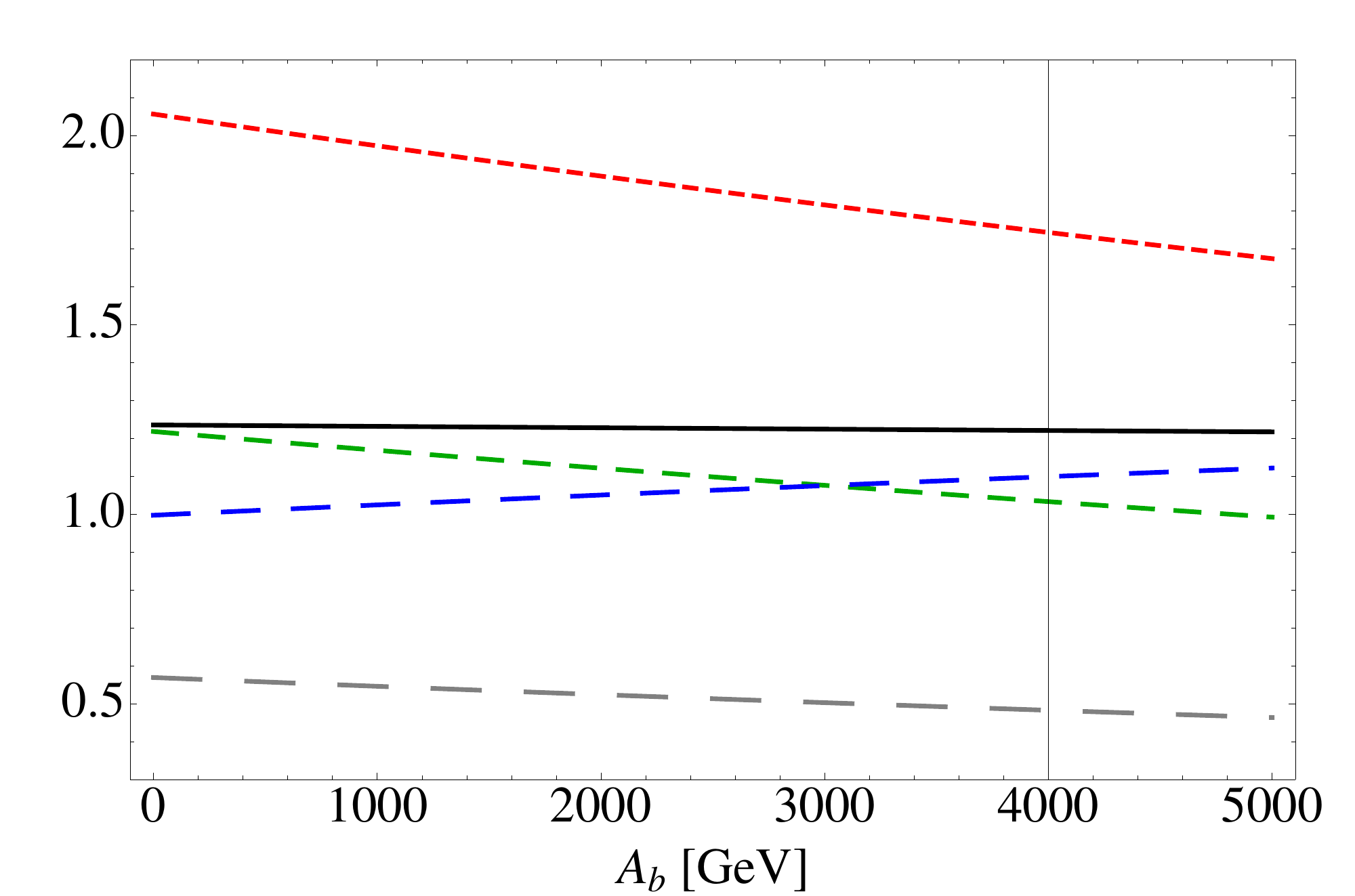}\hspace{-0.4cm}
\includegraphics[scale=0.37]{Figures/rates_legend}\hspace{-0.45cm}
\includegraphics[scale=0.35]{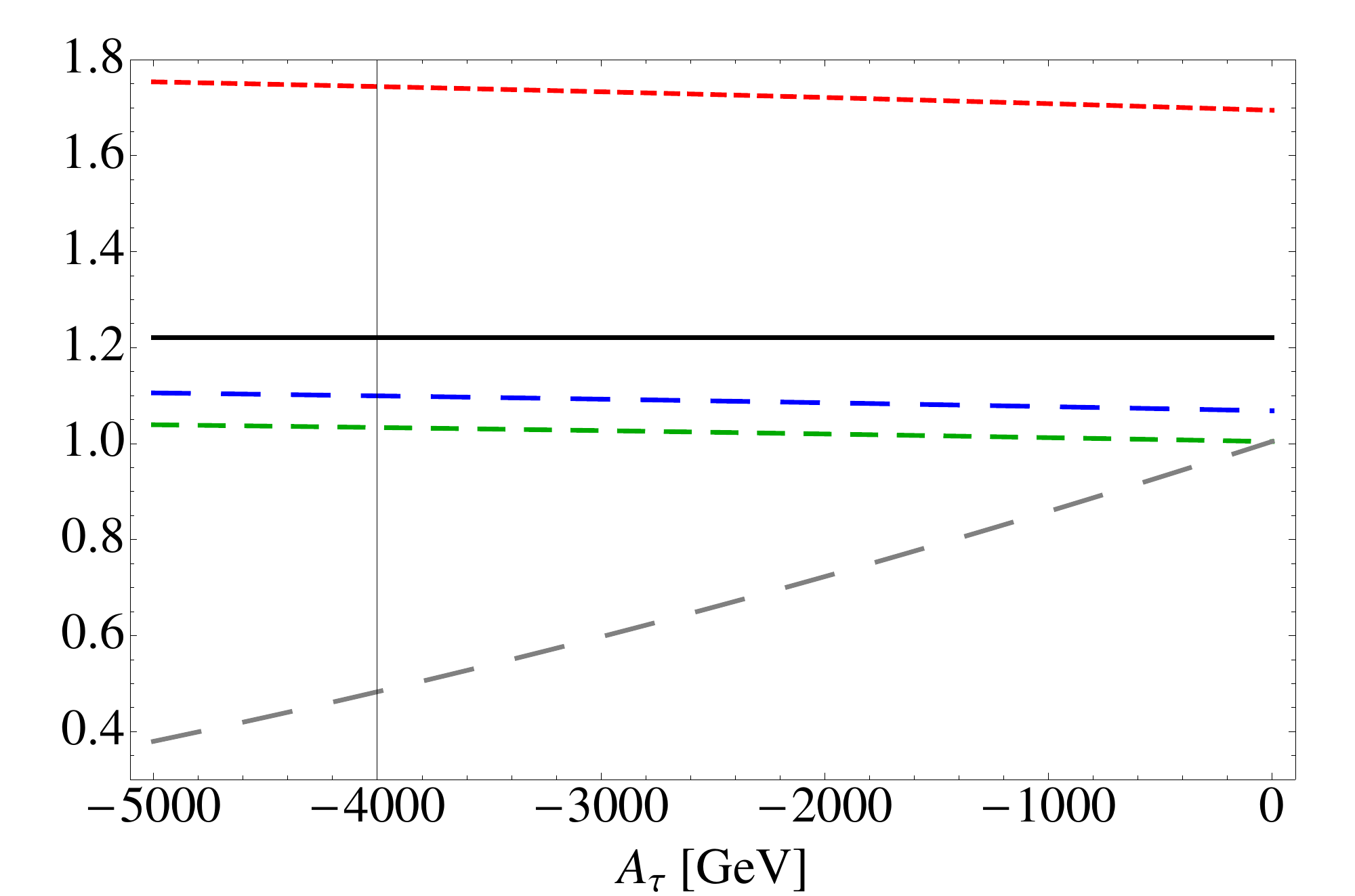}
\end{center}
\caption{ 
\small The signal rates and the coefficient $c_g^2$ as functions of the $A$-terms $A_b$ (left panel) and $A_\tau$ (right panel). In each case we keep fixed all the other parameters at their values given in Table \ref{bench}. The vertical lines correspond to the benchmark point values.
}\l{fig:Aterm}
\end{figure*}
\begin{figure*}[!t]
\begin{center}
\includegraphics[scale=0.4]{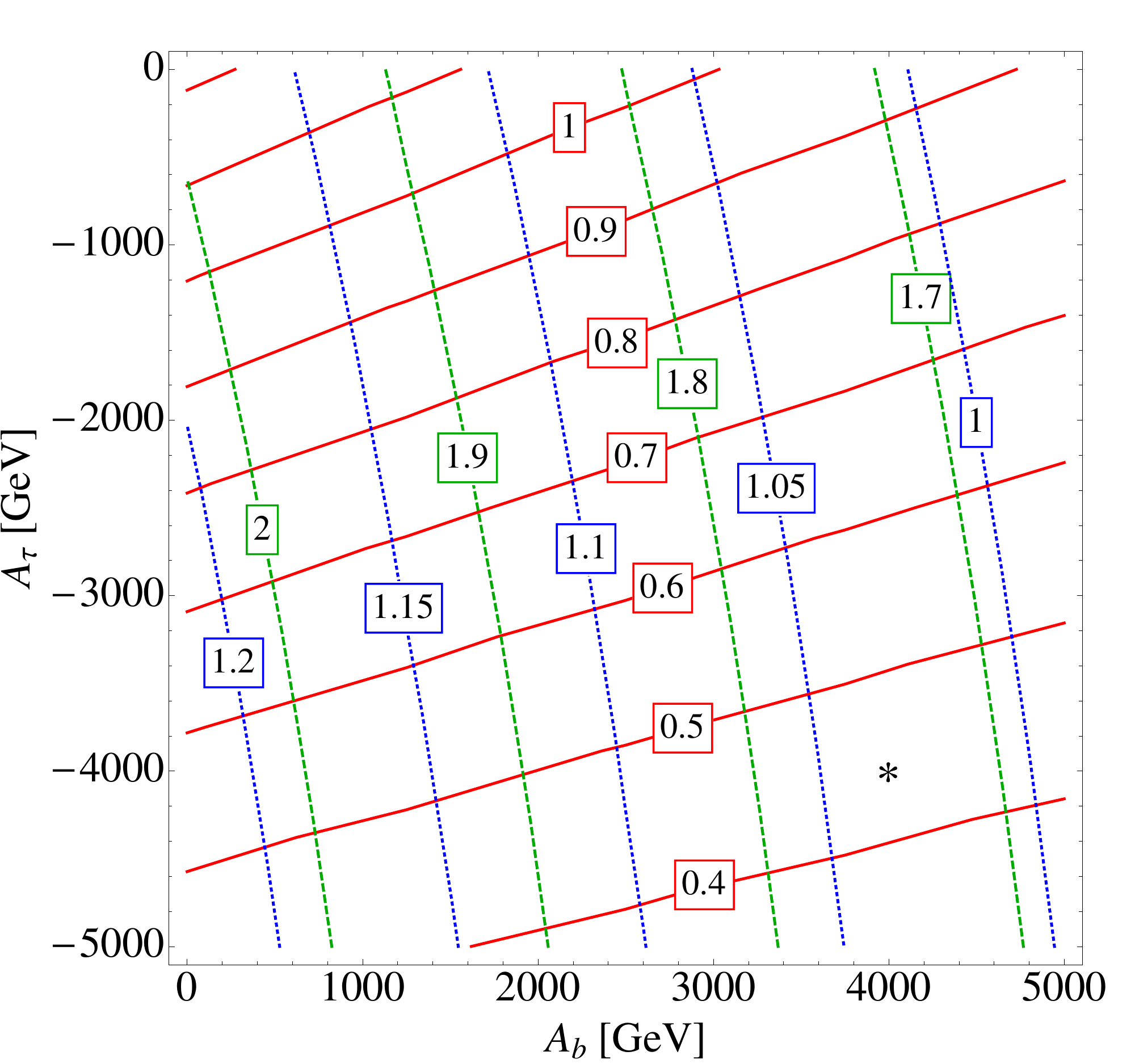}
\end{center}
\caption{ 
\small Contour plot showing the ratio $R^{\mathrm{incl}}_{\tau\tau} /R^{\mathrm{VH}}_{bb} $(red solid line), $R^{\mathrm{incl}}_{\gamma\gamma}$ (green dashed line) and $R^{\mathrm{incl}}_{ZZ^\ast}=R^{\mathrm{incl}}_{WW^\ast}$ (blue dotted line), as functions of the $A$-terms $A_b$ and $A_\tau$, keeping fixed all the other parameters at their values given in Table \ref{bench}. The dot $*$ represents the benchmark point.
}\l{fig:contour}
\end{figure*}

\begin{figure*}
\begin{center}
\hspace{-0.35cm}\includegraphics[scale=0.35]{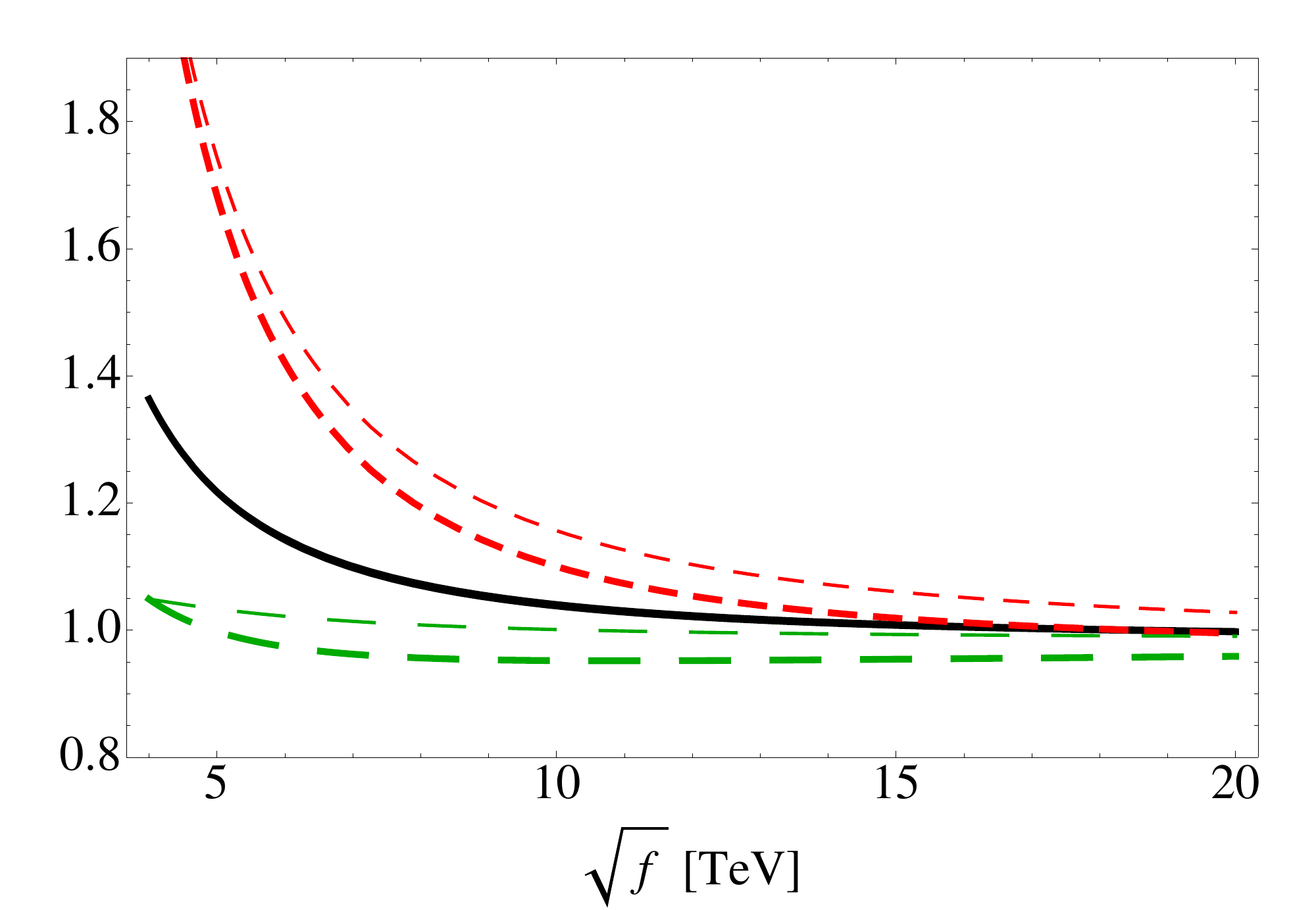}\hspace{-0.4cm}
\includegraphics[scale=0.37]{Figures/rates_legend}\hspace{-0.45cm}
\includegraphics[scale=0.35]{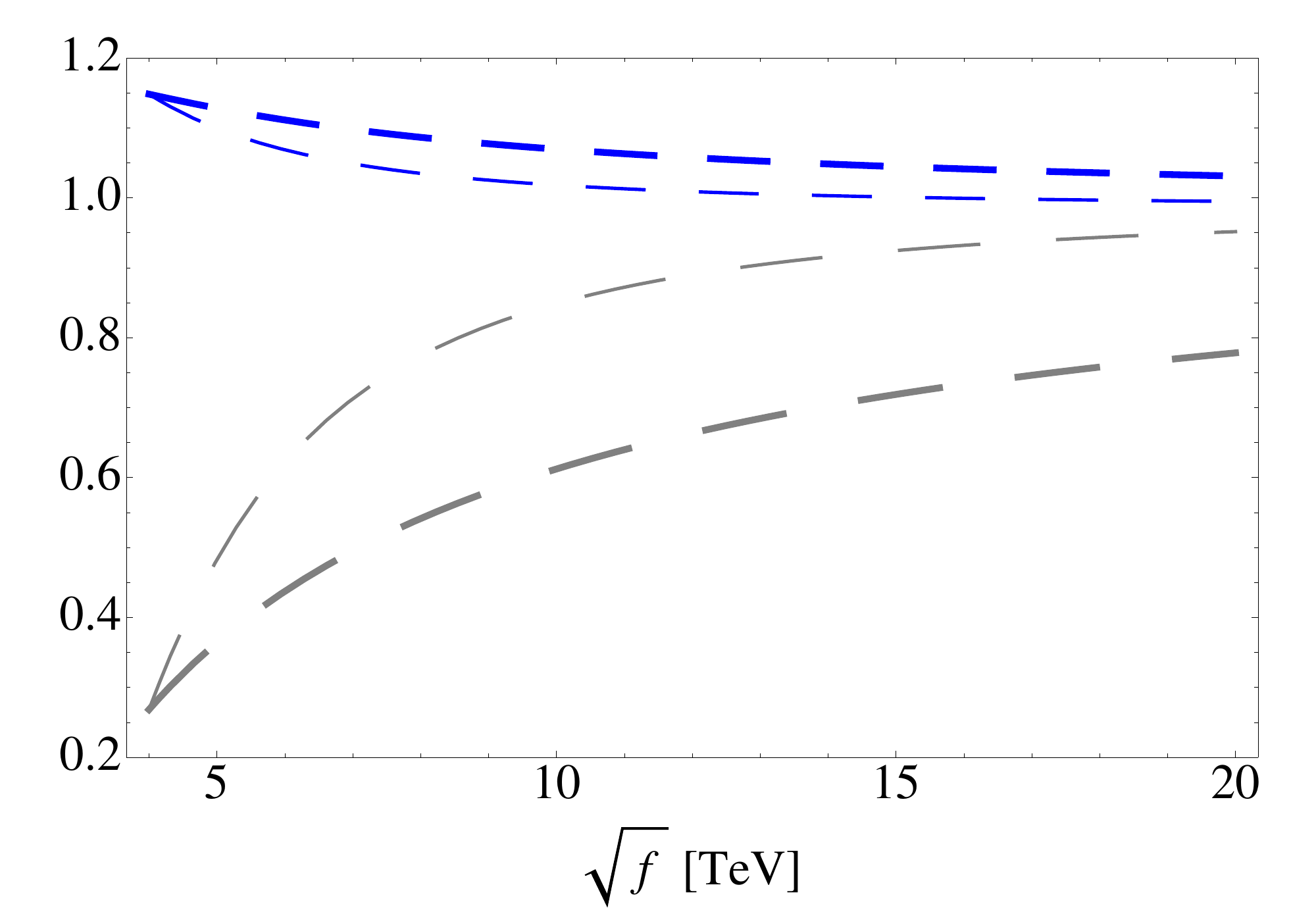}
\end{center}
\caption{ 
\small The coefficient $c_g^2$ and the signal rates $R^{\mathrm{incl}}_{\gamma\gamma}$, $R^{\mathrm{incl}}_{ZZ^\ast}=R^{\mathrm{incl}}_{WW^\ast}$ (left panel), $R^{\mathrm{incl}}_{\tau\tau}$ and $R^{\mathrm{VH}}_{bb} $(right panel), as functions of the SUSY breaking scale $\sqrt{f}$ with \mbox{$A_b=-A_\tau=4$ TeV} (thin lines) and $A_b=-A_\tau=\sqrt{f}$ (thick lines). The two lines for $c_g^2$ coincide. In both cases all the other parameters are kept fixed as in Table \ref{bench}.}\l{fig:fdep}
\end{figure*}

\subsection{Bounds from di-tau searches} \l{sgoldbound}

The bounds on the sgoldstino-like scalar, arising from di-photon and di-jet searches have been discussed in Ref.~\cite{Bellazzini:2012ul}. The key difference in the scenario considered in this paper is due to the presence of large $A$-term parameters, which, in particular, induce large couplings of the sgoldstino to the bottom quarks and the tau leptons and hence large BRs for the sgoldstino decays into $b\bar{b}$ and $\tau^+\tau^-$. Due to the fact that gluon-gluon fusion is the dominant production mechanism, the sgoldstino decay into $b\bar{b}$ is overwhelmed by the QCD background and can not be used to set bounds on the parameter space. However, the $\tau^+\tau^-$ final state provides an interesting channel from the point of view of both constraint and discovery prospects.

In Ref.~\cite{ATLASCollaboration:2012un} the ATLAS Collaboration recently presented a search for a high mass resonance decaying into $\tau^+\tau^-$ with an integrated luminosity of $4.6$ fb$^{-1}$ at a center of mass energy of $7$ TeV. This search focuses on neutral vector resonances ($Z'$) and unfortunately it cannot be directly used to set bounds on our model. However, in order to have an idea of the sensitivity of current searches of resonances in the $\tau^+\tau^-$ final state in terms of the scenario we are considering, we assume the selection efficiency and kinematic acceptance of the $Z'$ to be equal to the one of a neutral scalar. By this assumption, we can simply use the bounds presented by the ATLAS search in order to constrain the parameter space of our scenario. The exact structure of all the sgoldstino interactions is strongly model-dependent, so in order to be conservative, and with motivation from the fact that a large coupling of the sgoldstino to tau leptons is required in order to significantly affect the $c_\tau$-coupling in Eq.~\eqref{ct}, we assume BR$\(\phi\to \tau^+\tau^-\)$=1. This is a reasonable assumption in the case where $|A_{\tau}|$ is  larger than all the other soft parameters and where the mixing angle $R_{(h,x)}$ is small. In this case the bound of Ref.~\cite{ATLASCollaboration:2012un} can be directly compared to the total sgoldstino production cross section, which is plotted in Fig.~2 of Ref.~\cite{Bellazzini:2012ul}. In particular, by taking into account the sgoldstino production cross section via gluon-gluon fusion, given by,
\be\l{prodCS}
\bry{lll}
\dst \sigma_{\phi}&=&\dst \f{\pi}{32}\f{m_{3}^{2} \,m^{2}_{\phi}}{s f^{2}}\int_{m_{\phi}^{2}/s}^{1}\f{dx}{x}f_{p/g}\(x,m_{\phi}^{2}\)f_{p,\bar{p}/g}\(\f{m_{\phi}^{2}}{xs},m_{\phi}^{2}\)\,,
\ery
\ee
we can use the ATLAS search in order to obtain a $95\%$ CL exclusion region in the plane $\(m_{\phi},m_{3}/f\)$. This is done in Fig.~\ref{exclusion}, where we also indicate our benchmark point with a star. We have chosen our benchmark point very close to the experimental bound in order to show how large the deviation in $h\to\tau^+\tau^-$ can be in the scenario we are considering. However, as it is clear from Fig.~\ref{fig:fdep}, a larger value of $\sqrt{f}$ would bring us below the present bound.

The bound shown in Fig.~\ref{fig:fdep} suggests the following considerations. Since the ratio $m_{3}/f$ is minimized for low values of $m_{3}$, we choose the lowest value of $m_{3}$ that is compatible with the experimental lower limit on the gluino mass, at around $ 1$ TeV. The maximum value for $f$, for which we still can have an effect in the $h\to \tau^+\tau^-$ channel, can be read off the right panel of Fig.~\ref{fig:fdep}. If we expect a depletion of the order of $50\%$, then $\sqrt{f}$ cannot be much larger than $10$ TeV, even if we take $A_{\tau}$ to be at the edge of the perturbative region, i.e.~$A_{\tau}\sim\sqrt{f}$. Therefore, assuming \mbox{$\sqrt{f_{\text{max}}}=10$ TeV}, we can conclude that if the depletion in $h\to \tau^+\tau^-$ is confirmed by the LHC Collaborations and if indeed this can be explained by a mixing of the Higgs boson with a sgoldstino scalar, then we should expect $m_{3}/f\gtrsim 0.01$ TeV$^{-1}$, and due to the large value of $A_\tau$, a significative BR$\(\phi\to \tau^+\tau^-\)$. We can estimate the sensitivity to the $\phi\to \tau^+\tau^-$ process of the LHC with a higher integrated luminosity $L$. The simplest way of doing this is by the purely statistical assumption that the limit in Ref.~\cite{ATLASCollaboration:2012un} scales as $\sqrt{L}$. By this assumption, we see that with about one hundred times the luminosity used in the analysis of Ref.~\cite{ATLASCollaboration:2012un} we expect that the relevant $m_{3}/f$ region can be explored, allowing for a  possible confirmation or exclusion of our explanation for the $h\to \tau^+\tau^-$ depletion. This is of course a conservative estimation, since the LHC center of mass energy is going to increase and the understanding of the relevant systematic uncertainties is expected to improve. Therefore, new searches for resonances in the di-tau final state are strongly motivated, especially in the case of scalar ones.

\begin{figure*}[!t]
\begin{center}
\includegraphics[scale=0.42]{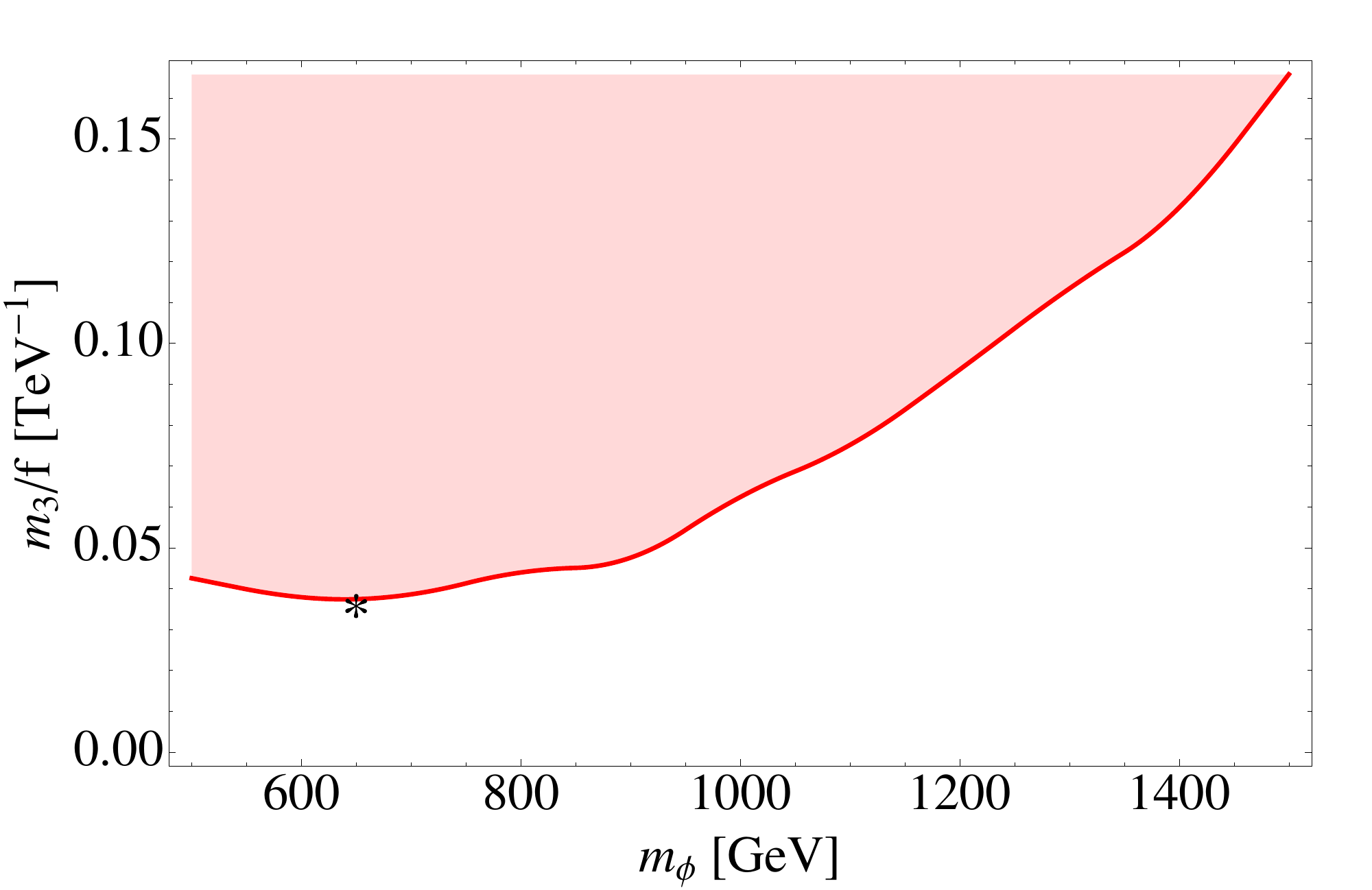}
\end{center}
\caption{ 
\small The region above the red line corresponds to the $95\%$ CL upper limit on $\sigma\(pp\to Z'\)\times$ BR$\(Z'\to \tau^+\tau^-\)$ of the ATLAS search of Ref.~\cite{ATLASCollaboration:2012un}. The exclusion in the plane $\(m_{\phi},m_{3}/f\)$ was obtained assuming that the experimental exclusion can be straightforwardly applied to a scalar resonance (i.e.~the kinematic acceptance and the selection efficiency is the same as for a $Z'$) and that BR$\(\phi \to \tau^+\tau^-\)=1$. The dot $*$ represents the benchmark point.
}\l{exclusion}
\end{figure*}

\section{Conclusions}\l{conclusion} 

In this paper we discussed the structure of the Higgs couplings that arise in a scenario where supersymmetry is broken spontaneously at a low scale. 
By promoting the MSSM soft terms to supersymmetic operators and treating the goldstino superfield dynamically, 
we showed how the mixing with the sgoldstino-like scalar can induce a dependence on the MSSM soft parameters for the couplings of the lightest Higgs scalar. In particular we discussed how the tree level Higgs couplings to fermions depend on the $A$-term parameters for the corresponding fermion. This dependence allowed for  modifications of the usual MSSM Yukawa couplings. In an attempt to isolate this effect from mixing effects between the two Higgs doublets we focused on the decoupling limit, in which the MSSM Yukawa couplings reduce to the SM ones. As an illustrative example, we discussed a benchmark model in which the Higgs coupling to tau leptons is depleted and the coupling to photons is enhanced, while all other couplings, including the one to bottom quarks, are close to their corresponding SM value. Of course, by refraining from taking the decoupling limit, a wider range of possibilities opens up. We also discussed the exclusion/discovery prospects for the sgoldstino-like scalar in terms of di-tau searches. In particular, we argued that a significant deviation in the Higgs coupling to tau leptons motivates searches for a scalar resonance in the di-tau final state.

\vspace{0.5cm}
\section*{Acknowledgments}

We thank Brando Bellazzini, Lorenzo Calibbi, Adam Falkowski and Chiara Rovelli for discussions. 
This work was partly supported by the MICINN under the grants FPA2009-07908 and FPA2010-17747.
The work of C.P. is supported in part by IISN-Belgium (conventions 4.4511.06, 4.4505.86 and 4.4514.08), by the ``Communaut\'e Fran\c{c}aise de Belgique" through the ARC program and by a ``Mandat d'Impulsion Scientifique" of the F.R.S.-FNRS. The work of A.R.~was partly supported by the Spanish MICINNs Juan de la Cierva and Consolider-Ingenio 2010 programme under grants CPAN CSD2007-00042, MultiDark CSD2009-00064, the Community of Madrid under grant HEPHACOS S2009/ESP-1473 and the European Union under the Marie Curie- ITN programme PITN-GA-2009-237920. The work R.T. is supported in part by the Research Executive Agency (REA) of the European Union under the Grant Agreement number PITN-GA-2010-264564 (LHCPhenoNet) and by the ERC Advanced Grant no.267985, Electroweak Symmetry Breaking, Flavour and Dark Matter: One Solution for Three Mysteries, (DaMeSyFla).


\bibliographystyle{JHEP}
\bibliography{draft}{}

\end{document}